\documentclass[aps,prl,twocolumn,groupedaddress,amsmath,amssymb,showpacs]{revtex4}
\usepackage[dvips]{graphicx}
\begin{document}

\title{   Thermal compression of two-dimensional atomic hydrogen to quantum degeneracy }

\author{S. Vasilyev,$^1$ J. J{\"a}rvinen,$^1$  A.I. Safonov,$^2$
 and S. Jaakkola$^1$}

\affiliation { $^1$Wihuri Physical Laboratory, Department of Physics, University of Turku, 20014 Turku, Finland\\
$^2$Laboratory of Metastable Quantum Systems, ISSSP, RRC Kurchatov
Institute, 123182 Moscow, Russia\\}

\email{servas@utu.fi}

\date{\today}

\begin{abstract}
We describe experiments where 2D atomic hydrogen gas is compressed
thermally at a small "cold spot" on the surface of superfluid
helium and detected directly with electron-spin resonance. We
reach  surface densities up to $5\times 10^{12}$ cm$^{-2}$ at
temperatures $\approx 100$ mK corresponding to the maximum 2D
phase-space density $\approx1.5$. By independent measurements of
the surface density and its decay rate we make the first direct
determination of the three-body recombination rate constant and
get the upper bound $L_{3b} \lesssim 2\times10^{-25}$ cm$^4$/s
which is an order of magnitude smaller than previously reported
experimental results.
\end{abstract}
 \maketitle

When adsorbed on the surface of superfluid helium spin-polarized
atomic hydrogen (H$\downarrow$) is an ideal realization of a
two-dimensional (2D) boson gas \cite{Walraven}. Helium provides a
translationally invariant substrate and its surface-normal
potential supports only one bound state for hydrogen with the
binding energy $E_a$=1.14(1) K \cite{OurEa}. Even for such a weak
interaction, lowering the surface temperature $T_s$ well below 1 K
leads to a large adsorbate density $\sigma$. At high H$\downarrow$
coverages three-body recombination is expected to be the dominant
density decay mechanism setting the main obstacle to the
achievement of the quantum degeneracy regime, where the thermal de
Broglie wavelength $\Lambda$ is larger than the average
interatomic spacing. Degenerate 2D H$\downarrow$ is expected to
exhibit collective phenomena such as the Kosterlitz-Thouless
superfluidity transition and the formation of a quasicondensate.

Two methods of local compression of adsorbed H$\downarrow$ have
been employed to overcome limitations caused by recombination and
its heat. Magnetic compression has been successfully used to
achieve quantum degeneracy \cite{OurMagComp,Mosk}. In this method
the recombination heat is removed from the compressed
H$\downarrow$ by ripplons of the helium surface and the cooling
efficiency depends on the length of the heat transfer path. By
decreasing the size of the compressed region to 20 $\mu$m
 we were able to achieve $\sigma\Lambda^2\approx9$ \cite{OurMagComp}. However, the small
size of the sample together with large magnetic field gradients
did not allow to implement direct diagnostics of adsorbed
H$\downarrow$. In the thermal compression method \cite{Svistunov,
Matsubara} cooling a small part of the sample cell wall well below
the temperature of the rest of the wall leads to an exponential
increase of $\sigma$ on such a "cold spot". In this method the
recombination heat is transferred from the ripplons to the phonons
of helium \cite{Settija} and then to the substrate beneath the
spot. Therefore a larger spot is preferable as long as the total
recombination rate on the spot becomes a limitation. The larger
sample size and the homogeneity of the magnetic field make thermal
compression better suited for direct studies of adsorbed
H$\downarrow$.

In the present work we use sensitive electron-spin resonance
\cite{Instability} to diagnose 2D H$\downarrow$ gas thermally
compressed to $\sigma\Lambda^2\approx1.5$ and discuss the
limitations and possible improvements of the "cold spot" method to
reach and detect the Kosterlitz-Thouless transition. By
independent measurements of the recombination rate and the surface
density we obtain the first directly determined value, actually an
upper bound, of the three-body surface recombination rate constant
$L_{3b}\lesssim 2\times 10^{-25}$ cm$^4$/sec. This is 4...10 times
smaller than the earlier values measured indirectly
\cite{OurMagComp, Bell, Reynolds, Sprik}.

\begin{figure}
\includegraphics[width=7.0cm]{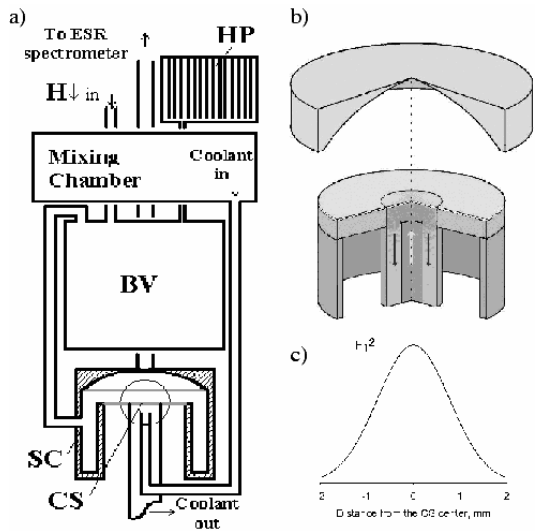}
\caption{(a) Scheme of the cell. SC = sample cell, CS = cold spot,
BV = buffer volume, HP = helium purifier; (b) Fabry-Perot
resonator with the cold spot in the center of its flat mirror; (c)
Microwave field profile on the cold spot.} \label{cell}
\end{figure}

Our experimental setup is shown in Fig.\ref{cell}. The sample cell
has been described elsewhere \cite{Instability}. The
low-temperature part of the ESR spectrometer, operating now as a
mm-wave bridge, has been modified so that we may use three times
smaller excitation powers and thus avoid ESR instability effects
\cite{Instability} in a wider temperatures range. Gradient coils
were built to reduce the field inhomogeneity of the main magnet so
that the width of the bulk line is decreased to $\approx0.1$ G. It
has been found that even very small ($\leq1$ ppb) $^3$He impurity
in our isotopically purified $^4$He condensed into the cell may
significantly influence the adsorption of the 2D hydrogen
\cite{OurEa}. Therefore we built an \textit{in situ} helium film
purifier, a reservoir with the large surface area of $2\times10^3$
cm$^2$ located in zero field (Fig.\ref{cell}). Being cooled down
to the lowest temperatures attained here the purifier adsorbs the
reminder of $^3$He atoms from the cell. The miniature RuO$_2$
thermometers used to measure the temperatures of the cell, the
buffer volume and the $^3$He/$^4$He coolant of the cold spot were
calibrated to an accuracy better than 1 mK with a $^3$He melting
curve thermometer.

In a typical experiment we fill the cell and the buffer with
atomic hydrogen from a low-temperature dissociator to a density
$n\approx10^{15}$ cm$^{-3}$. After switching off the discharge we
stabilize the buffer to $T_b=350$ mK, the optimum temperature for
getting the longest lifetime for the hydrogen sample, and cool the
cell to a desired temperature $T_c$=60...170 mK. The temperatures
of the mixing chamber of the dilution refrigerator and the
$^3$He-$^4$He stream cooling the cold spot (cp. Fig.\ref{cell})
are stabilized in the range $T_L$=30...150 mK. Thermal compression
can be rapidly ($<$ 5 s) turned on and off by changing the coolant
temperature. Decays of our H$\downarrow$ samples were governed by
impurity relaxation from "pure" hyperfine state $b$ to more
reactive state $a$ \cite{BlueBible}. To decrease the relaxation
rate the cell is made of nonmagnetic materials (ultrapure copper,
epoxy, Kapton foils) and its walls are coated at low temperatures
with a 10...50 nm thick solid H$_2$ layer by running the
dissociator for several days. The 0.2 cm diameter tube connecting
the buffer and the cell is wide enough to render the exchange of
atoms between the two volumes faster than the decay rates. This
ensures a dynamic density equilibrium between the volumes. Decays
of H$\downarrow$ samples were monitored at fixed temperatures of
the cell region. Recombination rates in the buffer and cell are
measured calorimetrically from the feedback powers of the
respective temperature controllers. We find that at the present
temperatures the loss rate of atoms in the buffer is negligible
compared to that in the cell. Integrating the latter we extract
the bulk densities as functions of time in both volumes.

\begin{figure}
\includegraphics[width=6cm]{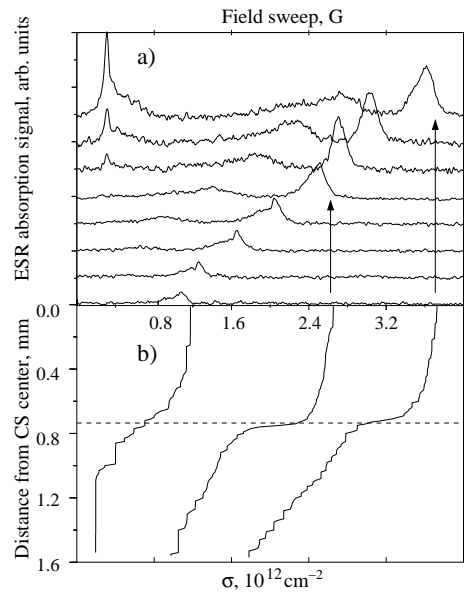}
\caption{Evolution of (a) H$\downarrow$ ESR spectrum and (b)
surface density profile during the decay of a H$\downarrow$
sample. The traces were recorded at intervals of 500 s at $T_L=45$
mK and $T_c=112$ mK. The dashed line marks the edge of the cold
spot.} \label{spectra}
\end{figure}

The evolution of the ESR spectrum during the decay of a
H$\downarrow$ sample is shown in Fig. \ref{spectra}. The resonance
line originating from the adsorbed atoms is shifted from the bulk
line due to the non-zero average dipolar field in this 2D system
\cite{Instability, Shinkoda91}. To find the surface density we
integrate the surface and bulk absorption ESR lines. The bulk
integral is calibrated calorimetrically against the absolute value
of the bulk density, which method does not rely on the adsorption
isotherm and yields the absolute value of surface density with an
accuracy of $10\%$. With this method we checked the relation
between the internal dipolar field and the surface density and
found $\Delta H_d = A \sigma$, where $A=1.05(10) \times 10^{-12}$
G cm$^2$. This direct and more accurate measurement agrees well
with the previous experimental and calculated results
\cite{Instability, Shinkoda91}.

The ESR lineshape of adsorbed atoms $S(h)$ as a function of
magnetic field sweep $h$ is broadened due to the inhomogeneity of
the temperature $T_s(r)$ and, consequently, density $\sigma(r)$ of
the 2D gas in the spot region. It is given by the relation
\begin{equation}
S(h) \sim \int{H_1^2(r) f_i(h-A\sigma(r)) \sigma(r) r dr}, \label{Lineshape}
\end{equation}
where $H_1^2(r)$ is the microwave field profile on the flat
mirror. The intrinsic lineshape $f_i(h)$ of the adsorbed atoms is
very narrow \cite{Instability} and can be replaced by a delta
function. We used Eq. (\ref{Lineshape}) to extract density
profiles $\sigma(r)$ from the observed lineshapes using a
numerical fitting routine. Examples of the surface density
profiles recovered for a few ESR spectra is presented in Fig.
\ref{spectra}b. Even for the largest temperature difference
between the cell and spot the surface density is homogeneous
within $10\%$ on the cold spot. The relatively slow decrease of
$\sigma$ outside the spot ($r>0.75$ cm) gives rise to a broad
maximum between the bulk and main surface signals.

\begin{figure}
\includegraphics[width=8 cm]{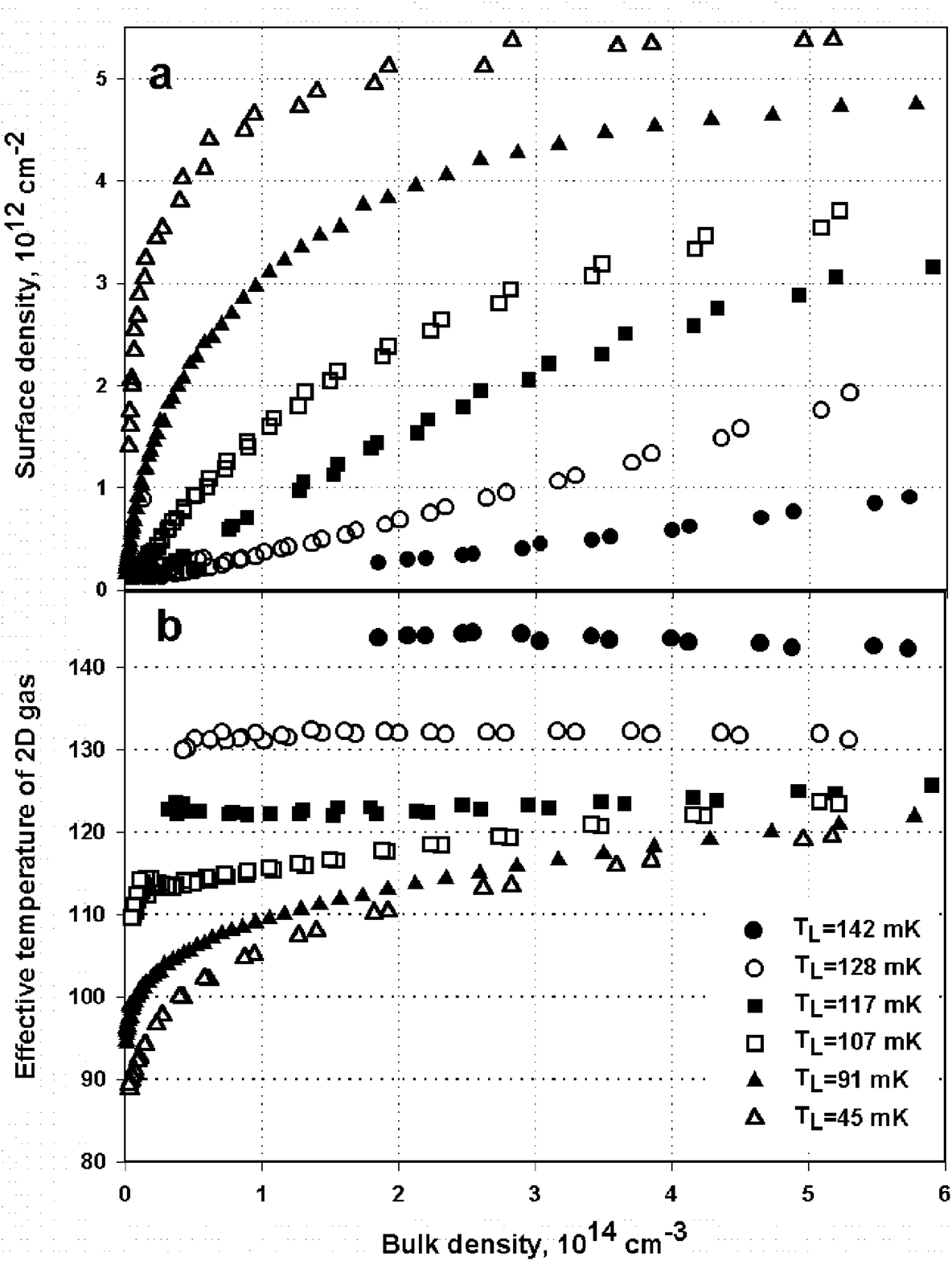}
\caption{(a) Maximum surface density on the cold spot for
$T_{C}$=154 mK and various $T_L$. (b) Effective surface
temperatures $T_s$ calculated using adsorption isotherm.}
\label{sigma}
\end{figure}

Numerous decays have been measured at various cell and spot
temperatures. Plots of the maximum surface density as a function
of the bulk density $n$ for a fixed $T_{c}$=154 mK and various
coolant temperatures $T_{L}$ are presented in Fig. \ref{sigma}a.
For low bulk densities the surface density increases linearly with
$n$, as it should do according to the adsorption isotherm
\cite{Svistunov, Moskthesis}. Effective temperature of the 2D gas
$T_s$ (Fig. \ref{sigma}b) is extracted from the $\sigma$ and $n$
values using the adsorption isotherm and taking $E_a$=1.14(1) K
\cite{OurEa}. At high temperatures $T_L\gtrsim100$ mK and small
differences $T_L-T_c\lesssim10$ mK, $T_s$ turns out be equal to
$T_L$ within 1 mK, the error bar of our thermometry. This
coincidence is regarded as a confirmation of the above mentioned
adsorption energy value.

The observed levelling of $\sigma$ with increasing $n$ (upper
plots in Fig. \ref{sigma}) points to an overheating of the 2D
H$\downarrow$ gas which begins the earlier the lower is the
coolant temperature. The surface temperatures $T_s$ extracted from
the low-density parts of the curves start to exceed the coolant
temperature, the difference increasing with increasing $T_c-T_L$.
This may be explained by heating of the spot by a heat flux from
the much warmer cell walls.  Another possible reason for the
saturation of the surface density could be a 2D hydrodynamic flow
of H$\downarrow$ on the spot out of the high density region. If
the flow becomes fast enough, the balance between the adsorption
and desorption rates will be disturbed. Interaction of
H$\downarrow$ with surface quasiparticles like ripplons and $^3$He
impurities should impede the flow. To distinguish between the
roles of the overheating and the 2D flow we added $^3$He into the
cell varying the $^3$He surface density up to $10^{14}$ cm$^{-2}$.
This did not help to get any higher density for the adsorbed
H$\downarrow$. On the other hand, the maximum $\sigma$ appeared to
be very sensitive to the rate of one-body impurity relaxation in
the cell and buffer. Therefore we conclude that the recombination
overheating is more important  than the 2D flow and limits the
highest achieved densities. The $T_s$ values extracted from the
adsorption isotherm (Fig. \ref{sigma}) can be considered as good
upper limits to the real surface temperatures.

\begin{figure}
\includegraphics[width=7 cm]{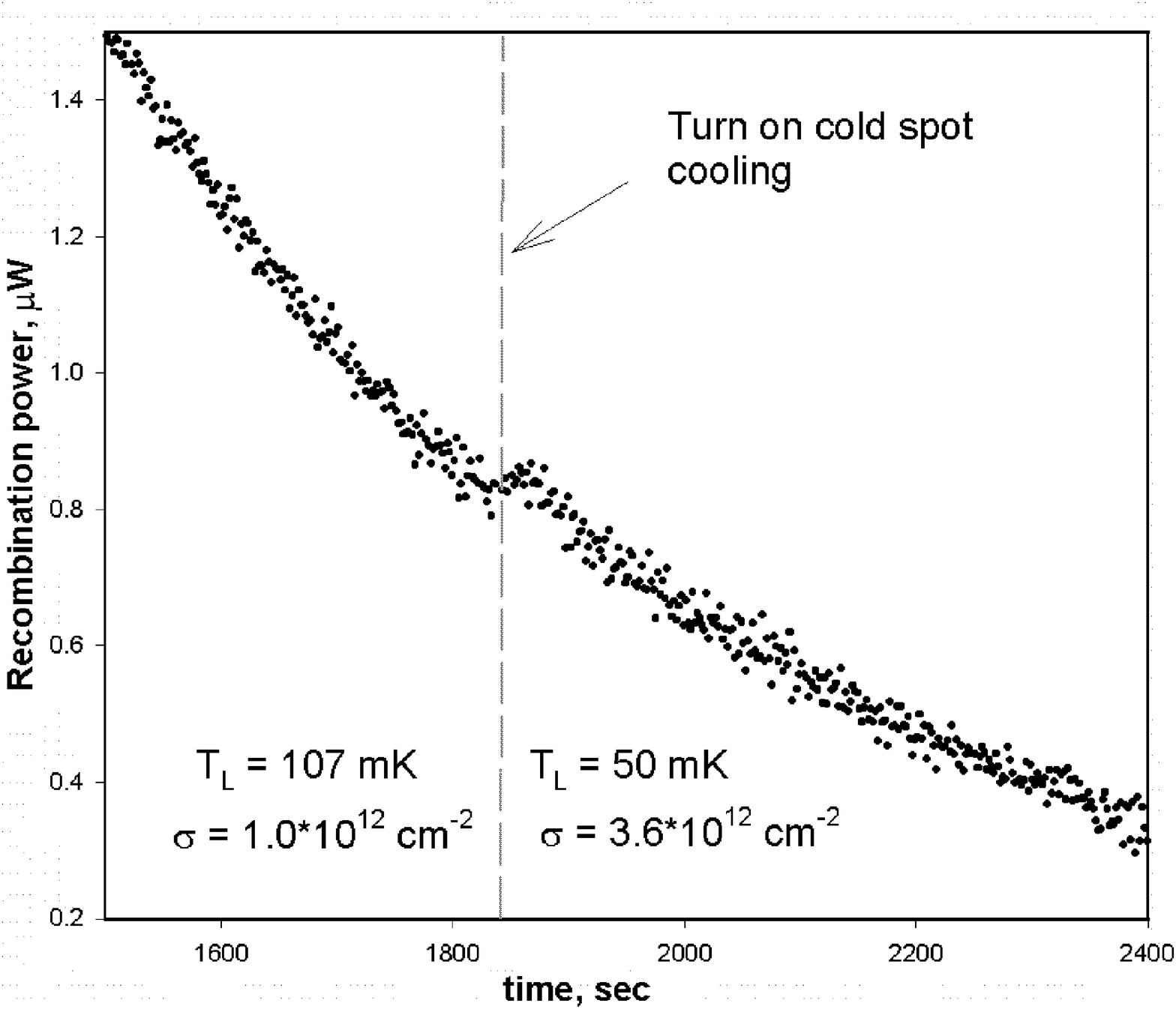}
\caption{Change in recombination power when the temperature of the
cold spot is rapidly decreased from 107 to 50 mK and the surface
density increases from $1\times10^{12}$ cm$^{-2}$ to
$3.6\times10^{12}$ cm$^{-2}$.} \label{recombination}
\end{figure}

In our experiments the decay of H$\downarrow$ is determined mainly
by one-body relaxation and three-body recombination on the cell
walls. Due to the density profile over the spot at high $\sigma$
being nearly step-like (Fig. \ref{spectra}) we can separate the
spot contribution and write for the total recombination rate
\begin{equation}
\frac{dN}{dt} = - (2G_1\sigma + L_{3b}\sigma^3)A_{cs} -
R(n,T_c,T_b), \label{TotalRate}
\end{equation}
where $G_1$ is the one-body relaxation rate constant, $L_{3b}$ the
three-body recombination rate constant, $A_{cs}$ the cold spot
surface area and $R(n,T_c,T_b)$ is the atom loss rate in the rest
of the cell and the buffer. We tried to extract the spot
contribution by monitoring recombination powers during decays at
different coolant temperatures and keeping conditions in the rest
of the cell region unchanged. No difference was found within the
noise of the temperature controller signal indicating that the
spot contribution is very small compared to that of the rest of
the walls. Therefore, to get a higher resolution, we used the
possibility to turn the spot cooling rapidly on (or off) during
the decay (Fig. \ref{recombination}) and performed such
experiments at various cell temperatures and bulk densities.
Even for the largest change in the surface density of the spot the
corresponding change of the total recombination power in the cell
was hardly discernible from the noise. Assuming that three-body
recombination is the dominant loss mechanism on the spot and
integrating the three-body rate over the actual density profile of
the spot we extract the upper limit $L_{3b} \leq 2\times10^{-25}$
cm$^4$/s. This is the first directly measured value based on
independent determinations of the recombination rate and surface
density. It agrees well with the calculations of de Goey et al.
\cite{deGoey}, but is an order of magnitude lower than the
$L_{3b}$ values reported in several previous works \cite{Bell,
Reynolds, Sprik}.

In all previous measurements of surface recombination $\sigma$ was
inferred into the rate equations through the adsorption isotherm.
Instead of Eq. (\ref{TotalRate}) a rate equation for the bulk
density
\begin{equation}
\frac{dn}{dt} = - G_1^{e}n - L_{3b}^{e}n^3, \label{DensityRate}
\end{equation}
was used. Here $G_1^{e}=G_1(A/V)\Lambda\exp(E_a/T_S)$ and
$L_{3b}^{e}=L_{3b}(A/V)\Lambda^3\exp(3E_a/T_S)$ are effective rate
constants with $A/V$ being the area-to-volume ratio of the cell.
The effective constants were extracted from the fits of the
measured $n(t)$ curves and the adsorption energy and the intrinsic
constants $G_1$ and $L_{3b}$ were obtained from the Arrhenius
plots of the effective constants. There are however two
fundamental drawbacks in this approach. First, $E_a$ and  the
intrinsic rate constants are correlated parameters, and even a
small uncertainty in $E_a$ can seriously affect the accuracy of
the intrinsic constants. A $^3$He content as small as 0.1 ppm can
be a reason for some of the published values of $E_a$ being
$\approx10\%$ lower than the latest value $E_a$ = 1.14(1) K
obtained for isotopically purified $^4$He \cite{OurEa} and
confirmed also in this work. At $T_s\approx100$ mK this error
decreases the factor $\exp(3E_a/T_s)$ and increases $L_{3b}$ by an
order of magnitude. Another problem arises from the uncertainty of
$T_s$ due to the recombination heating of the 2D gas. This makes
the surface temperature a function of time in the decays and does
not allow to use Eq. (\ref{DensityRate}) with $G_1^{e}$ and
$L_{3b}^{e}$ being constant. In the present method we do not rely
on the adsorption isotherm, but the error in $L_{3b}$ is
determined only by the absolute inaccuracy of $\sigma$ and is at
most 30\%.

Appearance of a superfluid 2D H$\downarrow$ on the cold spot is
expected to manifest itself as an abrupt change of the density
profile and ESR lineshape. In this work superfluidity was not yet
observed. The highest surface density
$\sigma\approx5\times10^{12}$ cm$^{-2}$ was achieved at
$T_s\approx100$ mK corresponding to the quantum degeneracy degree
of about 1.5. Lowering the temperature of the sample cell below
110 mK or increasing the bulk density above
$n\approx2\times10^{14}$ cm$^{-2}$ was found to excessively
increase the recombination rate on the cell walls  thus
overheating the 2D H$\downarrow$ on the cold spot. On the other
hand, when the cell walls were warmer than 150 mK we could not
cool the spot sufficiently due to a heat flux through the spot
substrate and thermal accomodation of the atoms from the bulk.
When designing the cell, three-body recombination on the spot was
thought to be faster than what it turned out to be and this was
the reason to limit the spot size. However, one-body surface
relaxation outside the spot was observed to give the dominant
contribution to the decay of the sample and appeared to be the
main limitation in the present experiments. We succeeded in
decreasing the rate of one-body relaxation to
$G_{1}\approx7\times10^{-2}$ s$^{-1}$ at $T_c\approx100$ mK,
smallest ever reported in H$\downarrow$ experiments \cite{Bell,
Reynolds, Sprik, Mosk}. Yet in future experiments one should find
a way to reduce the $G_1$ even further, e.g., by covering the cell
walls with a diamagnetic insulating material \cite{Moskthesis}.
Another modification to get higher surface densities and to make
the study of surface recombination more quantitative would be to
make the cold spot larger and thermally better insulated from the
cell. On the basis of the upper limit for $L_{3b}$ obtained in the
present work we estimate that such modifications would result in
an increase of the degeneracy parameter
$\sigma\Lambda^{2}\gtrsim3$, where the three-body recombination
probability starts to decrease due to quantum correlation effects
\cite{OurMagComp}.

\begin{acknowledgments}
We thank I.I. Lukashevich and A.A. Kharitonov for collaboration.
This work was supported by the Academy of Finland, Wihuri
Foundation, INTAS, the Russian Ministry of Industry, Science and
Technology and the RFBR.
\end{acknowledgments}

\bibliography{CS}

\end{document}